\documentclass[12pt,epsf]{article}

 \usepackage{epsfig}
 \usepackage{times}
 
 \usepackage{amsmath}
 \usepackage{amssymb}

\newcommand{\be}{\begin{equation}}
\newcommand{\ee}{\end{equation}}
\def\({\left (}
\def\){\right )}
\def\[{\left [}
\def\[{\right ]}

\begin{document}
\begin{titlepage}
\bigskip
\rightline{}
\rightline{hep-th/0610058}
\rightline
\bigskip\bigskip\bigskip\bigskip
\centerline {\Large \bf {Bubbles Unbound: Bubbles of Nothing Without Kaluza-Klein }}
\bigskip\bigskip
\bigskip\bigskip

\centerline{\large Keith Copsey}
\bigskip\bigskip
\centerline{\em Department of Physics, UCSB, Santa Barbara, CA 93106}
\centerline{\em keith@physics.ucsb.edu}
\bigskip\bigskip

\begin{abstract}
I present analytic time symmetric initial data for five dimensions describing ``bubbles of nothing'' which are asymptotically flat in the higher dimensional sense, i.e. there is no Kaluza-Klein circle asymptotically.   The mass and size of these bubbles may be chosen arbitrarily and in particular the solutions contain bubbles of any size which are arbitrarily light.   This suggests the solutions may be important phenomenologically and in particular I show that at low energy there are bubbles which expand outwards, suggesting a new possible instability in higher dimensions.  Further, one may find bubbles of any size where the only region of high curvature is confined to an arbitrarily small volume.

\end{abstract}
\end{titlepage}

\baselineskip=16pt
\setcounter{equation}{0}

\section{Introduction}

Kaluza-Klein ``bubbles of nothing'' were introduced a quarter of a century ago by Witten \cite{WittenBubbles} as an instability in the Kaluza-Klein (KK) vacuum.  By performing an analytic continuation on a Schwarzschild black hole he was able to find an instanton which describes the nucleation of a ``bubble'' where the Kaluza-Klein circle smoothly pinches off in the interior of the spacetime, resulting in a minimal two sphere.  Once produced the bubble accelerates out to null infinity, ``eating'' up the spacetime.  The production of these bubbles is fortunately forbidden in a theory with fundamental fermions and supersymmetric boundary conditions.   At the point where the circle pinches off (the end of the ``cigar'') the fermions are, by definition, antiperiodic.  Since the cigar is a simply connected manifold with a single spin structure a KK bubble requires antiperiodic boundary conditions for the fermions at infinity.  Those boundary conditions are, however, inconsistent with supersymmetry.  In the intervening years since their introduction bubbles have been useful in a wide variety of applications in time dependent spacetimes and black hole physics (see e.g. \cite{Elvangetal}  and \cite{HorowitzMaeda} and references therein).

One might wonder whether it is possible to find purely gravitational bubbles in asymptotically flat space.   Any such smooth solution presumably requires at least five dimensions and the present discussion will be limited almost entirely to that case.  Initially one might be skeptical that such solutions could exist; to form a bubble one needs a circle to pinch off and the asymptotic $S^3$ has none available.  Indeed it has become commonplace in the literature to refer to bubbles which asymptotically approach a flat Kaluza-Klein metric as asymptoticaly flat.  However, the symmetries, if any, of the interior of a spacetime need not be the same as its asymptotic symmetries.  The manifolds $S^1 \times S^2$ and $S^3$ are cobordant, so one only has to inquire whether this transition can occur in vacuum general relativity.   Fortunately, it is easy to answer in the afirmative; black rings \cite{EmpReall} smoothly interpolate between an asymptotic $S^3$ and a $S^1 \times S^2$ horizon.   To find the desired solution then one only need  find a metric where instead of the $S^1$ going to a finite value (the ``size'' of the black ring) it smoothly goes to zero.  

If such bubbles did exist they would likely be very interesting.  Such solutions would not have any inherent scale, unlike in the KK case, and hence one could find bubbles of either arbitrary size or positive mass.  Further the existence of such bubbles could not be ruled out by the supersymmetric boundary conditions argument given by Witten; the circle associated with the bubble will be absent by the time one reaches infinity.

Various cousins of this desired solution have appeared previously in the literature. LeBrun pointed out quite some time ago \cite{LeBrun} that it was possible to find negative mass bubbles in a spacetime which was locally, but not globally, asymptotically flat.  Much more recently Ross has found \cite{RossAdSBubbles} positive mass bubbles which are asymptotically AdS, although these solutions required a positive charge and depend crucially on the existence of a Chern-Simons term.    Bena, Warner and Wang have constructed \cite{BWW} asymptotically flat solutions with a large number of two-cycles supported by flux.

By considering an ansatz along the lines of the black ring solutions I construct time-symmetric initial data describing regular solutions which locally look like KK-bubbles but which are asymptotically flat.  The next section describes the form of these solutions and their geometric properties.   None of the solutions are static and the third section discusses the time evolution of the initial data for small times.  In particular I show there are small mass bubbles of arbitrary size which initially expand.  Hence, they represent a new possible instability of asymptotically flat space.  In the fourth section I examine the curvature of the solutions and note that for light solutions the curvature is small almost everywhere.  Finally, I summarize the solutions, discuss some of their implications and describe various possible generalizations and directions of research.

 \section{A New Bubble}
 
 \subsection{General Solution}
 
Motivated by form of black ring \cite{BRs}  and C metric solutions consider time symmetric initial data for a five dimensional solution with a $U(1) \times U(1)$ symmetry which is diagonal and apart from a typical overall conformal factor is factorizable:
\be \label{anstz}
ds^2 = \frac{R^2}{(x-y)^2} \Big[ A(x) \Big[-G(y) d\psi^2 - \frac{F(y)}{G(y)} dy^2 \Big] + B(y) \Big[H(x) d\phi^2 + \frac{J(x)}{H(x)} dx^2 \Big] \Big]
\ee
Note the constant R, which I take to have dimensions of length, could be absorbed into the various functions, but writing the metric this way allows the remaining parameters to be dimensionless.   Searching for  a solution for time symmetric initial data, one needs only satisfy the vacuum constraint (${} ^{(d - 1)}R = 0$).    This, combined with the factorized form of the solution, specifies the metric almost uniquely.   Consider asymptotically flat solutions containing a bubble formed by the $\psi$ angle pinching off, leaving a minimal two sphere parametrized by $x$ and $\phi$.  This implies that $g_{\psi \psi}$ vanishes at two values of y (at the bubble and at some point in the asymptotic $S^3$) and that $g_{\phi \phi}$ vanishes at two values of x (i.e. the poles of the minimal $S^2$).  Requiring that in such solutions where $g_{\psi \psi}$ and $g_{\phi \phi}$ vanish the metric is smooth gives the metric
\be \label{sol1}
ds^2 = \frac{R^2}{(x-y)^2} \Big[ A(x) \Big[-\frac{P(y)}{B(y)} d\psi^2 - \frac{B(y)}{P(y)} dy^2 \Big] + B(y) \Big[\frac{P(x)}{A(x)} d\phi^2 + \frac{A(x)}{P(x)} dx^2 \Big] \Big]
 \ee
 where
 \be
 P(\xi) = Q (\xi_4 - \xi) (\xi - \xi_3) (1 - \xi_2 \xi) (1- \xi_5 \xi)
 \ee
\be
A(x) = k_1 (1 - k_2 x)^2
\ee
and
\be
B(y) = k_3 (1- k_4 y)^2 
\ee
If one took $k_2 = k_4 = 0$ (\ref{sol1}) would just be the Euclidean charged C-metric.   That solution is automatically valid initial data in five dimensions since the four dimensional scalar curvature is proportional to the trace of the stress energy tensor of a Maxwell field (which vanishes in four dimensions).  However, that metric has conical singularities.   The parameters $k_2$ and $k_4$ allow one to eliminate these singularities.  Note regularity also requires that $A(x) \neq 0$ and $B(y) \neq 0$ for $x$ and $y$ in the allowed ranges; fortunately this requirement turns out to be consistent with the absence of conical singularities.

Spatial infinity is reached as both $x$ and $y$ go to $\xi_3$.  There is a bubble at the lower bound of $y$ where the coordinate $\psi$ pinches off.  This leads to, as desired, a minimal two sphere parametrized by $x$ and $\phi$.  This metric also contains a second bubble; $\phi$ pinches off at the upper bound of $x$ leading to a minimal two-sphere parametrized by $y$ and $\psi$.  Regularity of (\ref{sol1}) will not allow both x and y to have arbitrarily large ranges and, via a simple renaming of coordinates if necessary, one can always take x to be finite.  If $y$ also has a finite range then, without loss of generality, one may take
\be
1) \frac{1}{\xi_2} \leq y \leq \xi_3 \leq x \leq \xi_4 \, \, \, \, \, (\xi_2 \neq 0)
\ee
Requiring that the metric does not change signature implies the remaining zero of $P(\xi)$ is outside of the relevant ranges and leads to a series of subfamilies, i.e.
\be
a) \xi_5 = 0
\ee
\be
b) \frac{1}{\xi_5} <  \frac{1}{\xi_2}
\ee
\be
c) \xi_4  <  \frac{1}{\xi_5}
\ee
Alternatively, allowing the range of y to be semi-infinite
\be
2) -\infty < y \leq \xi_3 \leq x \leq \xi_4 < \frac{1}{\xi_5}
\ee
in which case regularity implies $k_4 \neq 0$, $\xi_2 = 0$ and $\xi_5 \neq 0$.  Treating case 2 carefully produces just the same results as those obtained from 1c in the limit $\xi_2 \rightarrow 0^{-}$ and for the sake of brevity I will state results explicitly only for case 1.   This somewhat exotic coordinate system is shown in Figure 1.  Note the two bubbles touch at a single point but the solution is perfectly regular there; locally the point is just the intersection of two orthogonal planes and as one moves away from the origin a plane wraps around each of the bubbles.
\begin{figure}
    \centering

	\includegraphics[scale= 0.7]{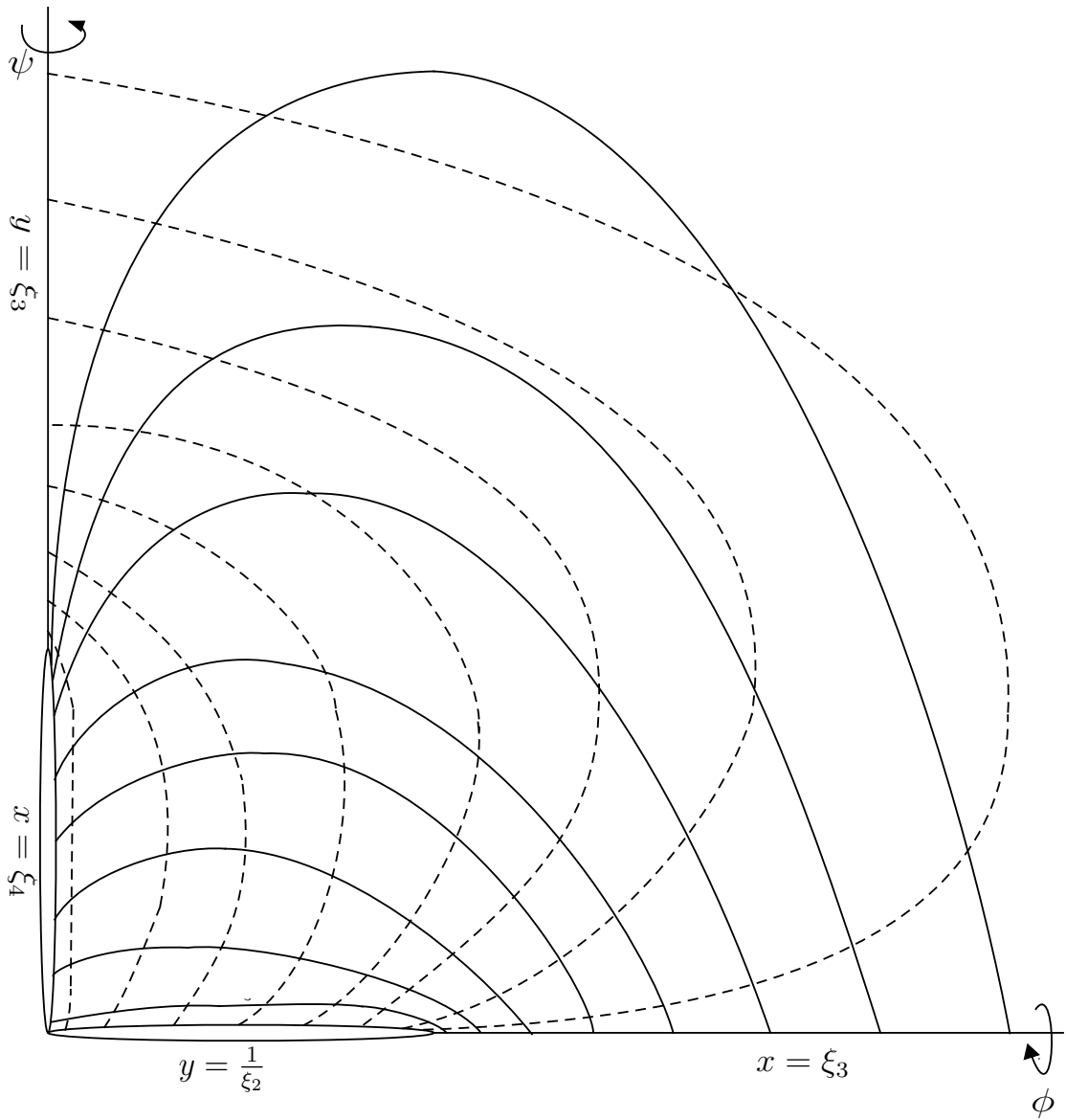}
	\caption{Lines of constant y (solid) and constant x (dashed).  $y = \xi_3$ and $x =  \xi_3$ are axis for $\psi$ and $\phi$, respectively, and bubbles are formed at $y= \frac{1}{\xi_2}$ as $\psi$ pinches off and at $x = \xi_4$  as $\phi$ pinches off.}
	
	\end{figure}

Demanding the absence of a conical singularity as $x \rightarrow \xi_4$ sets the period of $\phi$:
\be \label{phicon}
\Big \vert \frac{P'(\xi_4)}{2 A(\xi_4)} \Big \vert \Delta \phi = 2 \pi
\ee
Similarly, regularity as $y \rightarrow \frac{1}{\xi_2}$ sets the period of $\psi$:
\be \label{psicona}
\Big \vert \frac{P'(\frac{1}{\xi_2})}{2 B(\frac{1}{\xi_2})} \Big \vert \Delta \psi = 2 \pi
\ee
Later I will show that $k_2$ and $k_4$ can be chosen so that there are no conical singularities at $x =\xi_3$ or $y = \xi_3$ and doing so determines their values uniquely.

\subsection{Simplifying the Solution}

The solution (\ref{sol1}) simplifies substantially if one rewrites its parameters and variables in terms of physical quantities.  In particular, one can write the overall scale of the solution in terms of the size of one of the bubbles, say the one at constant y.  I will refer to this minimal $S^2$ as the y-bubble and define its size, $r_0$, via its area $A = 4 \pi r_0^2$.   Then one finds:
\be \label{scalarparams}
R^2  = r_0^2 \Big \vert \frac{Q (1 - \xi_2 \xi_3)(1- \xi_5 \xi_4)}{k_1 k_3} \Big \vert \frac{ (1-\xi_2 \xi_4)^2}{(1-k_2 \xi_4)^2 (k_4 - \xi_2)^2}
\ee
The area of the bubble at constant x, known hereafter as the x-bubble, is
\be
A' = \Delta \psi \int_{\frac{1}{\xi_2}}^{\xi_3} dy \sqrt{g_{y y}(x = \xi_4) g_{\psi \psi} (x = \xi_4)} =  4 \pi r_0^2 \Big \vert \frac{ (1 - \xi_5 \xi_4) (1 - \xi_2 \xi_3)}{(\xi_5 - \xi_2) (\xi_4 - \xi_3)} \Big \vert
\ee
It is handy to introduce a parameter $\omega$ ($0 < \omega < 1$) for the ratio of $A'$ to $A$:
\be
A' = 4 \pi {r_0}^2 \Big(\frac{1}{\omega} - 1 \Big)
\ee
Then $\omega \ll 1$ corresponds to $A'$ large relative to $A$ while $\omega \sim 1$ corresponds to an $A'$ small compared to $A$.  Equal size bubbles correspond to $\omega = 1/2$.  More generically, $A'$ is a monotonically decreasing function of $\omega$ and it is easy to write $\omega$ in terms of $A'/A$.   Specifically, defining $A' = 4\pi {r'_0}^2$,
\be
\omega = \frac{{r_0}^2}{{r'_0}^2 + {r_0}^2}
\ee
Note then
\be
1 - \omega = \frac{{r'_0}^2}{{r'_0}^2 + {r_0}^2}
\ee
and hence $\omega \leftrightarrow 1 - \omega$ if one exchanges the areas of the bubbles.

Now defining
\be
\bar{P}(\xi) =  (\xi_4 - \xi) (\xi - \xi_3) (1 - \xi_2 \xi) (1- \xi_5 \xi)
\ee
and  angles $\bar{\psi}$ and $\bar{\phi}$ such that the new angles have period $2\pi$, i.e.
\be
\bar{\phi} = \Big \vert \frac{P'(\xi_4)}{2 A(\xi_4)} \Big \vert \phi
\ee
and
\be
\bar{\psi} = \Big \vert \frac{P'(\frac{1}{\xi_2})}{2 B(\frac{1}{\xi_2})} \Big \vert \psi
\ee
and using  (\ref{scalarparams}) the metric (\ref{sol1}) can be written
\be \label{sol2}
ds^2 = r_0^2 \frac{(1-\xi_2 \xi_3)(1-\xi_5 \xi_4) (1- \xi_2 \xi_4)^2}{(1 - k_2 \xi_4)^2 (k_4 - \xi_2)^2 (x - y)^2} \Bigg\lbrace 
(1 - k_2 x)^2 (1 - k_4 y)^2 \Bigg[ \frac{dx^2}{\bar{P}(x)} - \frac{dy^2}{\bar{P}(y)} \Bigg] 
\ee
$$
- \frac{4 \Big(1 - \frac{k_4}{\xi_2} \Big)^4}{\Big(\bar{P}'(\frac{1}{\xi_2})\Big)^2} \frac{(1 - k_2 x)^2}{(1 - k_4 y)^2} \bar{P}(y) d \bar{\psi}^2 +  \frac{4(1 - k_2 \xi_4)^4}{\Big(\bar{P}'(\xi_4)\Big)^2}  \frac{(1 - k_4 y)^2}{(1 - k_2 x)^2} \bar{P}(x) d \bar{\phi}^2  \Bigg\rbrace 
$$
Note the scaling parameters (R, Q, $k_1$, $k_3$) have all vanished in favor of the size of the y-bubble $r_0$.

One can further shift and rescale the coordinates x and y:
\be \label{scaling}
\xi = \mathcal{A} \tilde{\xi} + \mathcal{B}
\ee
The zeroes of $P(\xi)$ automatically scale and shift appropriately (e.g. $\xi_4 = \mathcal{A} \tilde{\xi_4}+ \mathcal{B}$) but if the form of (\ref{sol2}) is to be unaltered under (\ref{scaling}) one must require that if $k_2 \neq 0$
\be
\frac{1}{k_2} = \mathcal{A} \frac{1}{\tilde{k}_2} + \mathcal{B}
\ee
and similarly if $k_4 \neq 0$
\be
\frac{1}{k_4} = \mathcal{A} \frac{1}{\tilde{k}_4} + \mathcal{B}
\ee
Provided this is true, then the only change of (\ref{sol2}) under (\ref{scaling}) is that the coordinates and parameters are replaced by the corresponding quantities with tildes.  These conditions turn out to be automatic once the absence of conical singularties is imposed.  Then $\mathcal{A}$ and $\mathcal{B}$ can be chosen to set one parameter $\xi_i$ to zero and rescale another to a desired constant.  In particular it will be convenient to take $\xi_3 = 0$ and $\xi_4 = 1$.  

Now turn to the choice of $k_2$ and $k_4$ such that conical singularities are absent.  To avoid a conical singularity  at $x = \xi_3$ take 
\be \label{k2}
k_2 = \frac{1 - \sqrt{\beta}}{\xi_4 - \sqrt{\beta} \xi_3}
\ee
where
\be
\beta = \Bigg \vert \frac{(1-\xi_2 \xi_4) (1 - \xi_5 \xi_4)}{(1-\xi_2 \xi_3) (1 - \xi_5 \xi_3)} \Bigg \vert
\ee
It is straightforward to check $\beta$ is invariant under scaling and shifting the zeroes $\xi_i$ and hence $k_2$ has the desired transformation properties.  The case $\sqrt{\beta} \xi_3 = \xi_4 $ can be dealt with either as a limit where $k_2 \rightarrow \infty$ or, noting if $k_2 \neq 0$ the metric (\ref{sol2}) can be written so only $1/{k_2}$ appears, as a point where $1/{k_2}$ (defined via the inverse of (\ref{k2}) ) vanishes.  In either case the metric is perfectly regular.\footnote{In this limit  $A(x) = k_1 (1-k_2 x)^2 \rightarrow  k_1 (k_2)^2 x^2$.  Since $\sqrt{\beta} \xi_3 = \xi_4 $ requires $\xi_3$ and $\xi_4$ to have the same sign, $x$, and thus $A(x)$, never vanishes.}  Also note any coordinate shift will take a divergent $k_2$ to a finite one:
\be
\frac{1}{k_2} = \mathcal{A} \frac{1}{\tilde{k}_2} + \mathcal{B}
\ee
In particular if one takes $\xi_3 = 0$, $k_2$ is always finite.  Finally, there is a second value of $k_2$ that would eliminate the conical singularity at $x = \xi_3$, but in that case $\xi_3 < 1/k_2 < \xi_4$ and so the metric would not be regular. 

Similarly to avoid a conical singularity at $y = \xi_3$ take
\be \label{k4}
k_4 = \frac{ \sqrt{\alpha} - 1}{\sqrt{\alpha} \xi_3 - \frac{1}{\xi_2}}
\ee
where
\be
\alpha = \Bigg \vert \frac{(\xi_4 - \frac{1}{\xi_2}) (1 - \frac{\xi_5}{\xi_2})}{(\xi_4- \xi_3) (1 - \xi_5 \xi_3)} \Bigg \vert
\ee
Making comments analogous to those above, $k_4$ has the desired transformation properties and $\sqrt{\alpha} \xi_2 \xi_3 =1$ is a perfectly regular limit.  This limit may again be avoided by shifting coordinates (in particular, such that $\xi_3 = 0$).  Finally, note one can check that while $k_2$ or $k_4$ may diverge individually they can not both do so simultaneously provided the zeroes $\xi_i$ are distinct.

Now consider generically whether $A(x)$ could have a zero in the range $\xi_3 \leq x \leq \xi_4$.   For $ 0 < \beta < 1$
\be
\frac{1}{k_2} - \xi_4 = \frac{\sqrt{\beta}(\xi_4 - \xi_3)}{1 - \sqrt{\beta}} > 0
\ee
and hence the zero of $A(x)$ is larger than $\xi_4$.  For $\beta = 1$, $k_2 = 0$ and $A(x) = 1$.  For $\beta > 1$
\be
\xi_3 - \frac{1}{k_2} = \frac{\xi_4 - \xi_3}{\sqrt{\beta} - 1} > 0
\ee
and  the zero of $A(x)$ is less than $\xi_3$.  For $\beta = 0$, $\frac{1}{k_2} = \xi_4 = \frac{1}{\xi_5}$ but this case is singular and excluded if the zeroes $\xi_i$ are distinct.

Similarly investigating whether $B(y)$ has a zero in the range $\frac{1}{\xi_2} \leq y \leq \xi_3$ if $0 < \alpha < 1$
\be
\frac{1}{\xi_2} - \frac{1}{k_4} = \frac{\sqrt{\alpha} \Big(\xi_3 - \frac{1}{\xi_2} \Big)}{1 - \sqrt{\alpha}} > 0
\ee
while for $\alpha > 1$
\be
\frac{1}{k_4} - \xi_3 = \frac{\xi_3 - \frac{1}{\xi_2}}{\sqrt{\alpha} - 1} > 0
\ee
and in both cases the zero of $B(y)$ lies outside of the relevant range.  For the case $\alpha = 1$, $k_4$ vanishes and there are no zeroes of $B(y)$.   If $\alpha = 0$, $k_4  = \xi_2$ would yield a singular case, but taking the $\xi_i$ to be distinct eliminates this possibility.  Finally note that while it is easiest to derive $k_2$ and $k_4$ away from spatial infinity ($x = y = \xi_3$) the given forms are sufficient to make the solutions asymptotically flat, as will be shown in detail below.

The forms of $k_2$ and $k_4$ mean one is free to go to a gauge where $\xi_3 = 0$ and $\xi_4 = 1$ and I now do so.  In fact, even after such a choice is made there is still gauge freedom left in (\ref{sol2}).  To see this define a physical coordinate $\bar{x}$ by the fraction of the area of the y-bubble covered in a disk from the pole at $x = 0$ to a given x:
\be
\bar{x} = \frac{2 \pi}{4\pi {r_0}^2} \int_{0}^{x} dx'  \sqrt{g_{x'x'} g_{\bar{\phi} \bar{\phi}}} = \frac{(1- \xi_2)x}{1 - \xi_2 x} 
\ee
and inverting
\be
x = \frac{\bar{x}}{ 1 - \xi_2 + \xi_2 \bar{x}}
\ee
Similarly, define a physical coordinate $\bar{y}$ via the area of the x-bubble between $y=0$ and a given y:
\be
1 - \bar{y} = \frac{2 \pi}{4\pi {r'_0}^2}  \int_{\frac{1}{\xi_2}}^{y} dy' \sqrt{g_{y'y'} g_{\bar{\psi} \bar{\psi}}} = \frac{1 - \xi_2 y}{1 - y}
\ee
or
\be
\bar{y} =  -\frac{(1 - \xi_2) y}{1 - y}
\ee
and inverting
\be
y = -\frac{\bar{y}}{1 - \bar{y} - \xi_2}
\ee
Note, by construction, $0 \leq \bar{x} \leq 1$, $0 \leq \bar{y} \leq 1$, there are bubbles at $\bar{x} = 1$ and at $\bar{y} = 1$ and spatial infinity is at $\bar{x} = \bar{y} = 0$.

Plugging this change of variables into the solution (\ref{sol2}) one finds the only parameters in the metric are $r_0$ and $\omega$:
\be \label{sol4}
ds^2 = \frac{{r_0}^2}{\omega (\bar{x} + \bar{y} - \bar{x} \bar{y})^2} \Big[ A(\bar{x}) \Big[\frac{4 P(\bar{y})}{B(\bar{y})} d{\bar{\psi}}^2 + \frac{B(\bar{y})}{P(\bar{y})} d{\bar{y}}^2 \Big] 
\ee
$$
+ B(\bar{y}) \Big[ \frac{A(\bar{x})}{R(\bar{x})} d{\bar{x}}^2 + \frac{4 R(\bar{x})}{A(\bar{x})} d\bar{\phi}^2 \Big] \Big]
$$
where
\be
A(\bar{x}) = \Big[ 1 - \Big(1 - \sqrt{1 - \omega}\Big) \bar{x} \Big]^2
\ee
\be
B(\bar{y}) = \Big[ 1 - \Big(1 - \sqrt{\omega}\Big) \bar{y} \Big]^2
\ee
\be
P(\bar{y}) = (1- \bar{y}) \bar{y} (1 - (1 - \omega) \bar{y})
\ee
and
\be
R(\bar{x}) = (1- \bar{x}) \bar{x} (1 - \omega \bar{x})
\ee
Note (\ref{sol4}) is determined entirely by the areas of the two bubbles.  In particular, then, (\ref{sol4}) should be invariant under exchanging the areas of the bubbles ($r_0 \leftrightarrow r'_0$).  Recalling this exchange implies $\omega \leftrightarrow 1 - \omega$ and noting that
\be
\frac{{r_0}^2}{\omega} = {r'_0}^2 + {r_0}^2
\ee
relabeling the coordinates $(\bar{x}, \bar{\phi}) \leftrightarrow (\bar{y}, \bar{\psi})$ demonstrates this symmetry.   On the other hand, (\ref{sol4}) also implies that one can determine the entire geometry by examining in detail one of the bubbles (e.g. $\omega$ is determined uniquely by $r_0$ and the minimal proper distance on the y-bubble between the poles $\bar{x} = 0$ and $\bar{x} = 1$).

\subsection{Geometric Properties}

In order to see the solution (\ref{sol4}) is indeed asymptotically flat and to obtain the quantities to  be used in calculating the mass it is handy to rewrite the metric in asymptotically spherical coordinates.  The simplest such coordinates which explicitly reflect the invariance under $r_0 \leftrightarrow r'_0$ and lead to a metric diagonal through leading order (i.e. $1/r^2$) corrections are
\be \label{newcoords}
\bar{x} = \frac{4 ({r_0}^2 + {r'_0}^2) \cos^2(\theta)}{r^2 + 4 {r'_0}^2 \cos^2(\theta) + 4{r_0}^2} \, \, \, \, \, \, \, \, \, \, \, \, \, \bar{y} = \frac{4 ({r_0}^2 + {r'_0}^2) \sin^2(\theta)}{r^2 + 4 {r_0}^2 \sin^2(\theta) + 4{r'_0}^2} 
\ee
Then the asymptotics of the metric become 
\be \label{metricasymptotics}
ds^2 = \Big[1 + \frac{\delta_{rr}}{r^2} + \mathcal{O}\Big(\frac{1}{r^4}\Big) \Big]dr^2 +  r^2 \Big[1 + \frac{\delta_{\theta \theta}}{r^2} + \mathcal{O}\Big(\frac{1}{r^4}\Big) \Big]d\theta^2
\ee
$$
  +  \mathcal{O}\Big(\frac{1}{r^3}\Big)dr d\theta + r^2 \sin^2(\theta) \Big[1 + \frac{\delta_{\bar{\psi} \bar{\psi}}}{r^2} + \mathcal{O}\Big(\frac{1}{r^4}\Big) \Big]d\bar{\psi}^2$$
$$+  r^2 \cos^2(\theta) \Big[1 + \frac{\delta_{\bar{\phi} \bar{\phi}}}{r^2} + \mathcal{O}\Big(\frac{1}{r^4}\Big) \Big]d\bar{\phi}^2$$
where
\be
\delta_{rr} = 4 (r_0 + r'_0)\bar{r} - 6 \bar{r}^2 + 2 (r_0  - r'_0)(r_0 + r'_0 - 2\bar{r}) \cos(2\theta)
\ee
\be
\delta_{\theta \theta} = 4 (r_0 + r'_0)\bar{r} - 2\bar{r}^2+ 2 (r_0  - r'_0)(r_0 + r'_0 - 2\bar{r}) \cos(2\theta)
\ee
\be
\delta_{\bar{\psi} \bar{\psi}} = 4 \Big({r_0}^2 + (r'_0 - r_0) \bar{r}  + \bar{r} (r_0 + r'_0 - \bar{r}) \cos(2\theta) \Big)
\ee
\be
\delta_{\bar{\phi} \bar{\phi}} =   4 \Big({r'_0}^2 + (r_0 - r'_0) \bar{r} + \bar{r} (\bar{r} - r_0 - r'_0) \cos(2\theta)\Big)
\ee
and $\bar{r} =  \sqrt{{r_0}^2 + {r'_0}^2}$.  Recall $\bar{\psi}$ and $\bar{\phi}$ both have period $2 \pi$ and so there are no conical singularities asymptotically.  It is fairly easy to see these asymptotics mean the solutions have finite ADM mass (the explicit expression are presented below) and are asymptotically flat under, for example, the definition of Mann and Marolf \cite{MannMarolf}.

Given a spacelike slice $\Sigma$, the ADM mass for asymptotically flat spaces may be written in covariant form as
\be
E = \frac{1}{16 \pi G} \int dS^{a} N \Big(D^{b}(\delta h_{a b}) - h^{b c} D_{a} (\delta h_{b c}) \Big)
\ee
where the integral is over over the (asymptotic) spatial boundary of $\Sigma$, $h_{a b}$ is the spatial metric induced on $\Sigma$ and $\delta h_{a b} = h_{a b} - h^{0}_{a b}$ where $h^{0}_{a b}$ is the spatial metric induced by a background (in this case flat) metric on $\Sigma$.  For asymptotic line elements of the form (\ref{metricasymptotics}) this gives
\be \label{Espherical}
E =  \frac{1}{16 \pi G} \int \Big(3 \delta_{rr} + \delta_{\theta \theta} + \delta_{\bar{\psi} \bar{\psi}} + \delta_{\bar{\phi} \bar{\phi}} \Big)
\ee
where the integral is over a unit three sphere.  For the metrics in question this means 
\be \label{Er}
E = \frac{2 \pi}{G} \sqrt{{r'_0}^2 + {r_0}^2} \Big[r_0 + r'_0 - \sqrt{{r'_0}^2 + {r_0}^2}  \Big]
\ee
While (\ref{Er}) displays explicitly the symmetry $r_0 \leftrightarrow r'_0$, the conditions for low and high energy solutions are somewhat complicated.  It is often simpler to regard the solutions as a function of $\omega$ and one of the bubble sizes, say $r_0$.  Then the energy can then be written as
\be
E = \frac{2 \pi r_0^2}{G \omega} \Big[ \sqrt{ 1- \omega} + \sqrt{\omega} - 1 \Big]
\ee
Noting the term in brackets is positive due to the triangle inequality, the energy is positive definite, as one expects from the positive energy theorems \cite{WSY}. Further, one can show that E is a monotonically decreasing function of $\omega$.  For $\omega \ll 1$
\be
E \approx \frac{\pi r_0^2}{G \sqrt{\omega}}
\ee
while for $\omega \sim 1$
\be
E \approx \frac{2 \pi r_0^2}{G} \sqrt{1 - \omega}
\ee
and so, given any positive energy and y-bubble size $r_0$, $\omega$ is determined.  Note for the sake of brevity I will refer to solutions where $E/{r_0}^2$ is large (small) as high (low) energy solutions.  

For the sake of concreteness I have chosen to use $r_0$ as a parameter and only if $\omega > 1/2$ does this correspond to the area of the larger bubble.  At the cost of some added complication one could, however, select the size of the larger  bubble and the energy independently:
\be \label{Egenbig}
E =  \frac{2 \pi {r_>}^2}{G} \sqrt{1 + \frac{{r_<}^2}{{r_>}^2}} \Bigg(1 +  \frac{r_<}{r_>} -  \sqrt{1 + \frac{{r_<}^2}{{r_>}^2}} \Bigg)
\ee
It is then straightforward to show (\ref{Egenbig}) is a monotonically increasing function of ${r_<}/{r_>}$.

Consider in detail the shape of the bubbles in the low energy limit ($r_0$ fixed, $\omega \sim 1$).  In this case it turns out the curvature of the y-bubble becomes vanishingly small away from the x-bubble.  Defining
\be \label{rhodef}
\rho = 2 r_0 \sqrt{\bar{x}}
\ee
the metric on the y-bubble becomes
\be \label{surfmet1}
ds^2 = \frac{ \Big( 1 + \sqrt{1 - \omega} \frac{\bar{x}}{1-\bar{x}}\Big)^2}{ \Big( 1 + (1 - \omega) \frac{\bar{x}}{1-\bar{x}}\Big)} d \rho^2 + \frac{\Big( 1 + (1 - \omega) \frac{\bar{x}}{1-\bar{x}}\Big) }{\Big( 1 + \sqrt{1 - \omega} \frac{\bar{x}}{1-\bar{x}}\Big)^2 } \rho^2 d \bar{\phi}^2 
\ee
and so provided $\sqrt{1 - \omega} \ll 1$, until $\bar{x}$ gets close to one
\be
1 - \bar{x} \sim  \sqrt{1 - \omega}
\ee
 the metric on the bubble can be approximated as flat space.   Physically what has happened is that all the curvature on the $S^2$ has been pushed into a small disk around the point $\bar{x} = 1$.  By definition this disk has a small proper area and examining (\ref{sol4}) for a moment it is clear the disk has a small proper radial distance (i.e. along the $\bar{x}$-direction).    The circumference around this region is still of order $r_0$ since the disk of high curvature is smoothly matched onto an almost flat region.  As a result there is large curvature on the y-bubble in the region around $\bar{x} \sim 1$.

On the other hand, in the low energy limit the x-bubble becomes an undistorted $S^2$.  Defining
\be
\sin(\theta) = 2 \sqrt{\bar{y} (1 - \bar{y})}
\ee
and taking $\theta(0) = 0$ the metric (\ref{sol4}) on the x-bubble becomes
\be
ds^2 = {r'_{0}}^2 \Bigg[ \frac{\Big[1 - (1 - \sqrt{\omega})\sin^2\Big(\frac{\theta}{2}\Big)\Big]^2}{1 -(1 - \omega) \sin^2\Big(\frac{\theta}{2}\Big)} d\theta^2 
\ee
$$
+\frac{1 -(1 - \omega) \sin^2\Big(\frac{\theta}{2}\Big)}{\Big[1 - (1 - \sqrt{\omega})\sin^2\Big(\frac{\theta}{2}\Big)\Big]^2} \sin^2(\theta) d \bar{\psi}^2 \Bigg]
$$
As $\omega \rightarrow 1$ with $r_0$ fixed, $r'_0$ becomes small and the x-bubble becomes an unsquashed small $S^2$.  While the x-bubble ``touches'' the y-bubble at $\bar{x} = 1, \bar{y} = 1$, the two bubbles have no directions in common and the large distortion on one does not affect the other.  As mentioned above, in the high energy limit the role of the bubbles will be reversed.  In other words, in the limit one bubble is much larger than the other the smaller one becomes an undistorted $S^2$ while the larger one becomes nearly flat everywhere except near the small bubble.

Finally one would like to know whether any of these solutions lie inside an apparent horizon and hence simply describe a slightly more exotic means of making a black hole than collapsing a shell of matter or gravitational radiation. Finding apparent horizons for these solutions analytically is technically difficult and I will leave detailed the consideration for numerical analysis.  One can, however, make a few relevant observations.  In the case one bubble is of fixed size and the other bubble is made arbitrarily small the energy, and hence the size of any event horizon, becomes arbitrarily small.  More generically, one can make a rough estimate of when the bubbles are inside an event horizon by comparing their energy to that of a KK black hole (or equivalently a black string of finite length) of Schwarzschild radius $r_s$.  There is of course no KK direction asymptotically for these bubbles but in the intermediate region between the bubble and infinity the proper distance $l$ around $\psi$ and $\phi$ is of order ${r_0}/\sqrt{\omega} = \sqrt{{r_0}^2 +{r'_0}^2}$.  Then
\be
M_{bs} = \frac{l r_s}{2 G_5}\sim \frac{r_s \sqrt{{r_0}^2 + {r'_0}^2} }{G_5} \sim \frac{\sqrt{{r_0}^2 + {r'_0}^2} \Big[r_0 + r'_0 - \sqrt{{r'_0}^2 + {r_0}^2}\Big]}{G_5}
\ee
and so
\be
r_s \sim r_0 + r'_0 - \sqrt{{r'_0}^2 + {r_0}^2}
\ee
Hence if one bubble is much smaller than the other the Schwarzschild radius of the black hole is of the same order as the small bubble size.  On the other hand, if the two bubbles are comparable in size the bubbles might be inside a horizon.  This suggests bubbles of comparable size will collapse and I now turn to such dynamical questions.

 \setcounter{equation}{0}

\section{Initial dynamical behavior}
I have described only the initial states for these bubbles.  In fact, an examination of the asymptotics is sufficient to show none of the solutions are static.  While the dynamical evolution of this initial data will hopefully be explored via numerical techniques, analytically one may at least discuss the initial time dependent behavior.  In particular, one would like to ask whether the bubbles initially expand or contract.  The technique of Corley and Jacobson \cite{CJ} requires spherical symmetry and hence does not help here.  However, one can simply use the Hamiltonian evolution equations in Gaussian coordinates (i.e. $N = 1, N^{a} = 0$).  For a vacuum solution of time symmetric initial data evaluated at the symmetry point (t = 0) the expressions are rather simple:
\be \label{timedeph}
\ddot{h}_{a b}(0) = \frac{ 2 N}{\beta \sqrt{h}} \Big( {\dot{\pi}^{G}}_{a b} + \frac{\dot{\pi}^G}{2 - d} h_{a b} \Big) =   \frac{ 2 N}{\beta \sqrt{h}}  \Big[ -\beta N \sqrt{h} \Big( {}^{(d - 1)}R_{a b} - \frac{ {}^{(d - 1)}R}{2} h_{a b} \Big)
\ee 
$$
+ \frac{h_{a b}}{2 - d} \Big( -\beta N \sqrt{h}\Big) \Big(1 - \frac{d}{2}\Big){}^{(d - 1)}R \Big] = -2 \, \, \,  {}^{(d - 1)}R_{a b} 
$$
where all expressions are evaluated at the moment of time symmetry.

One can then easily check that, at least through order $t^2$, the metrics for these bubbles retain their $U(1) \times U(1)$ symmetry and are diagonal with the exception of the developement of an $xy$ cross-term.\footnote{Specifically
$$
\ddot{h}_{\bar{x} \bar{y}} = \frac{ 4 (\sqrt{ 1- \omega} + \sqrt{\omega} - 1)}{(\bar{x} + \bar{y} -\bar{x} \bar{y}) (1 - \bar{x} + \bar{x} \sqrt{1 - \omega})(1 - \bar{y} + \bar{y} \sqrt{1 - \omega})}
$$
Hence, with the same observation about the triangle inequality as for the energy, $\ddot{h}_{\bar{x} \bar{y}} > 0$.} Further, any zeroes of an angle, say $\bar{\psi}$, remain at the same coordinates for any regular initial data since
\be
{}^{(d-1)}R_{\bar{\psi} \bar{\psi}} = {}^{(d-1)}R_{a b}\Big(\frac{\partial}{\partial  \bar{\psi}}\Big)^{a} \Big (\frac{\partial}{\partial  \bar{\psi}}\Big)^{b} = 0
\ee
evaluated at any point $ \bar{\psi}$ vanishes initially.  Then, since the area of the y-bubble is given (at least through order $t^2$) by
\be
A = 2\pi \int_{0}^{1} d\bar{x} \sqrt{g_{\bar{x} \bar{x}} \, g_{\bar{\phi} \bar{\phi}}}
\ee
where all quantities are evaluated at $\bar{y} = 1$,
\be
\dot{A} = 2\pi  \int_{0}^{1} d\bar{x} \frac{ \dot{g}_{\bar{x} \bar{x}} \, g_{\bar{\phi} \bar{\phi}} + g_{\bar{x} \bar{x}} \, \dot{g}_{\bar{\phi} \bar{\phi}}}{2 \sqrt{g_{\bar{x} \bar{x}} \, g_{\bar{\phi} \bar{\phi}}}}
\ee
and the initial acceleration of the area is
\be
\ddot{A} = 2 \pi  \int_{0}^{1} d\bar{x} \frac{ \ddot{g}_{\bar{x} \bar{x}} \, g_{\bar{\phi} \bar{\phi}} + g_{\bar{x} \bar{x} } \, \ddot{g}_{\bar{\phi} \bar{\phi}}}{2 \sqrt{g_{\bar{x} \bar{x} } \, g_{\bar{\phi} \bar{\phi}}}} 
\ee
$$
= - 2 \pi \int_{0}^{1} d\bar{x} \frac{ {}^{(d-1)}R_{\bar{x} \bar{x}} \, g_{\bar{\phi} \bar{\phi}} + g_{\bar{x} \bar{x} } \,  {}^{(d-1)}R_{\bar{\phi} \bar{\phi}}}{ \sqrt{g_{\bar{x} \bar{x} } \, g_{\bar{\phi} \bar{\phi}}}}
$$
evaluated at $\bar{y} = 1$.  The integral turns out to be an elementary one and yields
\be
\ddot{A} = \frac{8 \pi}{3 \sqrt{1 - \omega}} \Big[ 3 \sqrt{\omega} - 2 \sqrt{1 - \omega} - (1 + \omega) \Big]
\ee
The acceleration of the area is a dimensionless quantity and hence depends only on the ratio $r'_0/r_0$ or, equivalently, $\omega$.   In fact it is a monotonically increasing function of $\omega$.  For $\omega \ll 1$ (high energy solutions)
\be
\ddot{A} \sim -8 \pi
\ee
while for $\omega \sim 1$ (low energy solutions)
\be
\ddot{A} \sim \frac{8 \pi}{3\sqrt{1 - \omega}}
\ee
which is to say the y-bubble begins to expand at nearly the speed of light.

The transition between contraction and expansion occurs at
\be
\omega \approx 0.803815
\ee
corresponding to a relative ratio of the two bubble sizes
\be
\frac{r'_0}{r_0} \approx 0.494032
\ee
and energy
\be
E \approx  2.65366 r_0^2
\ee

The results for the acceleration of the x-bubble may be obtained simplify via $\omega \leftrightarrow 1 - \omega$.  For the sake of comparison, however, it is useful to write them out explicitly:
\be
\ddot{A}' = \frac{8 \pi}{3 \sqrt{\omega}} \Big[ 3 \sqrt{1 - \omega} + \omega - 2 - 2\sqrt{\omega} \Big]
\ee
and hence  $\ddot{A}'$ is a monotonically decreasing function of $\omega$.  For high energy solutions ($\omega \ll 1$)
\be
\ddot{A}'  \sim \frac{8 \pi}{3 \sqrt{\omega}} 
\ee
while for low energy solutions ($\omega \sim 1$)
\be
\ddot{A}'  \sim - 8 \pi
\ee
and the transition between expansion and contraction of $A'$ occurs at
\be
\omega \approx 0.196185
\ee
which translates into a relative ratio of the two bubble sizes
\be
\frac{r'_0}{r_0} \approx 2.02416
\ee
and an energy of
\be
E \approx 10.8726  r_0^2
\ee
Hence if the bubbles are of comparable size both collapse while if one is much larger than the other the larger bubble expands while the smaller one contracts.  Energetically this behavior does not seem surprising; the regions of large curvature tend to collapse while the regions of small curvature expand to relax away their gradient energy.  This type of behavior is also familiar from the Kaluza-Klein context, although of course in that context ``large'' and ``small'' are relative to the KK scale \cite{CJ} . 

In terms of stability, however, this result is rather disturbing.  In particular at low energy one can form a bubble of arbitrary size which begins expanding at nearly the speed of light.  If this expansion of the larger bubble continues for any significant period of time the spacetime far away from the initial disturbance is radically altered and the bubbles represent a stimulated instability.  While the results in this section do not answer the question of the full time evolution of the bubbles, it seems unlikely that a bubble which begins expanding will stop at some point and start contracting.   In the KK case, numerical studies of bubbles \cite{Sarbach} have found that bubbles which begin expanding continued to do so indefinitely.   The closest KK analogy to these bubbles are arguably ones in which the KK circle grows asymptotically instead of going to a fixed size.  In fact, such solutions have been constructed  \cite{OferEvaGary} by analytically continuing the Myers-Perry solutions \cite{MP} and in five or more dimensions these bubbles also expand forever.  The initial data I have described includes bubble of arbitrarily large size which, at low energy, expand so any process halting the expansion would have to be dependent entirely upon the dynamics.  Finally, as discussed above, this expansion seems to be driven energetically and there is no apparent reason why the process would reverse at some point. 

I have so far only discussed the dynamics of a single pair of bubbles.  Several pairs of bubbles should, however, be able to nucleate far away from each other.   This then introduces the possibility several pairs of bubbles could collide.  In the KK case \cite{HorowitzMaeda} this type of collision produces a spacelike singularity that extends out to null infinity.   The singularity has a horizon, but the spacetime resembles maximally extended Schwarzschild including the second asymptotic region and a white hole.  If a similar process takes place in the present context even the existence of bubbles for relatively short periods of time could have dramatic consequences for the spacetime.

Collapsing bubbles which are relatively heavy should form black holes which will then evaporate.  Note even in the event one bubble expands indefinitely, the smaller bubble apparently collapses; in this case one may well produce a small black hole in the throat of an expanding bubble.  A Kaluza-Klein versions of that system has been previously described \cite{EmpReall02} by Emparan and Reall. Then these solutions suggest small black holes may be relatively easy to produce even in models without large extra dimensions.  A black hole is, however, not necessarily the only result of such a collapsing bubble; some light bubbles might become sufficiently small that quantum corrections become significant over the surface of the bubble before a horizon can form.  If so then one would have a quantum version of cosmic censorship violation and quantum gravity effects could be accessible to distant observers.

 \setcounter{equation}{0}

\section{Large Curvatures and $\alpha'$ Corrections}

From a quantum mechanical, and in particular stringy, point of view one only expects a supergravity solution to be trustworthy when the curvature is less than the Planck scale.  Hence this section discusses the square of the Riemann tensor for these solutions.  It turns out for time symmetric initial data this is equivalent to any other measure of the curvature, in a sense which will be detailed below.  First one would like to write the square of the d-dimensional (in the present case d = 5) Riemann tensor in terms of the (d-1)-dimensional quantities derived from initial data.  Using the Gauss-Codacci relation
\be
{}^{(d-1)} {R_{a b c}}^{d} = h_{a}^{f} h_{b}^{g} h_{c}^{k} {h^{d}_{j}} \,  {}^{(d)} {R_{f g k}}^{j}  - K_{a c} K_{b}^{d} + K_{b c} K_{a}^{d}
\ee
where $h_{a b}$ is the metric induced on the spatial slice of the initial data and $K_{a b}$ is the extrinsic curvature of that slice.  Going to Gaussian normal coordinates (i.e. lapse N = 1 and shift $N^{a} = 0$) for time symmetric initial data ($K_{a b} = 0$) one finds the square of the Riemann tensor may be expressed
\be \label{Riemanngen}
{}^{(d)} R_{a b c d} {}^{(d)} R^{a b c d} = {}^{(d -1)} R_{a b c d} {}^{(d - 1)} R^{a b c d} + 4  \, {}^{(d -1)} R_{a b} {}^{(d - 1)} R^{a b}
\ee
Note that the quantities in (\ref{Riemanngen}) are contracted with their respective natural metrics, i.e. the spacetime metric $g_{a b}$ on the left hand side and the induced spatial metric $h_{a b}$ on the right.  In view of the fact that in Gaussian normal coordinates ${}^{(d - 1)} R_{a b} = -\ddot{h}_{a b}/2$ (\ref{timedeph}) one may view (\ref{Riemanngen}) as the result that the d-dimensional curvature is the sum of curvature due to spatial gradients and the curvature due to time dependence.  Note that the right hand side of (\ref{Riemanngen}) is a sum of positive definite terms and hence the only way it can become small is for the curvature to become small.  This demonstrates that for time symmetric initial data $R^2$ being small is entirelty equivalent to all components of the Riemann tensor being small (in good coordinates).  This is, of course, in contrast to the more generic situation where one may have null curvatures which leave $R^2$ small while allowing components of the Riemann tensor to become large.

For the particular solutions (\ref{sol1}) under consideration the square of the Riemann tensor is rather lengthy.  However, for low energy solutions ($\omega \sim 1$)
\be \label{Riemann2}
\frac{12 (\bar{x} + \bar{y} - \bar{x} \bar{y})^2} { {r_0}^4  (1 - \bar{x}+ \bar{x} \sqrt{1 - \omega})^{12}}  \Big[(1-\bar{x})^6  (1 - \omega)\Big({\bar{x}}^2 (3 {\bar{x}}^2 - 2 \bar{x} + 3) 
\ee
$$
+ 2 \bar{x} \bar{y} (3 {\bar{x}}^2 + 1)(1 - \bar{x}) + (1 - \bar{x})^2(3 \bar{x}^2 + 2 \bar{x} + 3) {\bar{y}}^2 \Big) + \mathcal{O}\Big( (1-\omega)^2 (1-\bar{x})^5 \Big) \Big]
$$
Then, away from $\bar{x} \sim 1$,  for fixed $r_0$ the square of the Riemann tensor becomes arbitrarily small as $\omega \rightarrow 1$.  Since $R^2$ is the sum of positive definite terms (\ref{Riemanngen}), this implies the Riemann tensor is arbitrarily small, component by component, in good coordinates.   In particular, recall from section 2 that the surface of the y-bubble away from the x-bubble approaches flat space.  Note finally the omitted terms in (\ref{Riemann2}) contain multiplicative factors of $\bar{x}$ and $\bar{y}$ I have not written explicitly so that the given term is dominant everywhere, including asymptotically, aside from near $\bar{x} \sim 1$.

Near the x-bubble, the metric is perfectly regular but curvatures become large.  For $\bar{x} \sim 1$ the square of the Riemann tensor is
\be \label{Riemannx}
\frac{1}{\Big(1 - \bar{x} + \bar{x} \sqrt{1 - \omega}  \Big)^{12} {r_0}^4} \Big[24 (1 - \omega)^4 + 96 (1-\omega)^{\frac{7}{2}} (1 - \bar{x})
\ee
$$
+ 192 (1 - \omega)^{3} (1 - \bar{x})^2  + 288 (1 - \omega)^{\frac{5}{2}} (1 - x)^3+ 312  (1 - \omega)^{2} (1 - x)^4 
$$
$$
+ 192 (1 - \omega)^{\frac{3}{2}} (1 - \bar{x})^5 + 48 (1 - \omega) (1 - \bar{x})^6 + \ldots \Big]
$$
where the omitted terms are suppressed relative to the given ones by a factors of $\sqrt{1 - \omega}$ and $1-\bar{x}$.  As a simple order of magnitude estimate it is easy to see (\ref{Riemannx}) is large only when $1 - \bar{x} \lesssim \sqrt{1 - \omega}$ and in this case
\be
R^2 = \mathcal{O}\Big(\frac{1}{(1 - \omega)^2 {r_0}^4 } \Big)
\ee
More precisely, one can check that the given terms in (\ref{Riemannx}) have no extrema for $\omega \sim 1$ and so $R^2$ is largest at the x-bubble, as one would expect physically, and has a maximum value of
\be
R^2 = \frac{24}{(1 - \omega)^2 {r_0}^4} + \mathcal{O}\Big(\frac{1}{(1 - \omega)^\frac{3}{2} {r_0}^4} \Big)
\ee

By construction the region of high curvature ($\bar{x} \lesssim \sqrt{1 - \omega}$) covers a small proper area of the y-bubble.  Of course, to be precise, once one has such a region of high curvature the y-bubble can only reliable be described as a minimal sphere outside of that area.  Further consider a surface of constant $\bar{x}$ ($\bar{x} < 1$) bounding this region of high curvature.  The area of this three surface is
\be
A_{\bar{x}} = 4 \pi^2 \int_0^{1} d\bar{y} \sqrt{g_{\bar{y} \bar{y}} g_{\bar{\psi} \bar{\psi}}g_{\bar{\phi} \bar{\phi}}}
\ee
$$
= \frac{8 \pi^2 {r_0}^3}{(\omega \bar{x})^\frac{3}{2}} (1 + \bar{x} \sqrt{\omega}) \sqrt{(1-\bar{x})(1 - \omega \bar{x})} ( 1 - \bar{x} + \bar{x} \sqrt{1 - \omega})
$$
and so for $\omega \sim 1$ and $\bar{x} \sim 1$ this volume is small.  In particular, for $\omega \sim 1$ if $1 - \bar{x} \sim \sqrt{1 - \omega}$
\be
 A_{\bar{x}} \sim 32 \pi^2 {r_0}^3 (1 - \omega)
 \ee
Hence taking the y-bubble size to be fixed but the energy small one obtains large curvature near the x-bubble but small curvature everywhere else.  Intuitively, this should not be surprising; spatial gradients require energy and as one takes the limit that the energy goes to zero any significant gradients must be confined to a vanishingly small volume.   Note this also implies that one can always confine $\alpha'$ effects to a region much smaller than the area of the larger bubble regardless of how small the later becomes.   Of course, should the smaller ``bubble'' become of order the string scale $\alpha'$ corrections will become significant in that region and the description of supergravity there will not be reliable.  Apart from this small region, however, curvatures will remain small in the rest of the solution and hence there supergravity should remain a reliable guide. It should be noted, however, other quantum effects are expected to be important even for the larger bubble if it reaches the string scale.  Of course, supergravity will only be a good guide everywhere provided the smaller bubble is always significantly larger than the string scale.

Regarding quantum corrections, one should note that the asymptotic properties of these solutions, and in particular the mass, should not be affected.   I have referred to these solutions as bubbles, but in the region where quantum corrections are large this has been a matter of terminology rather than a result.  Quantum corrections might potentially change the geometry in some fundamental way or more likely keep one from talking about geometry.  These corrections should not, however, make a classically smooth solution not regular.   On the contrary, one expects that string theory should smooth out many singularities and in particular produce smooth solutions from what classically are singular positive mass black holes.   It is hard to see how a theory could do this but make classically smooth but strong curvature regions such as those I have described singular.

\section{Summary and Discussion}

I have presented a two parameter family of asymptotically flat solutions describing pairs of bubbles where one can either choose the sizes of the bubbles or the size of one bubble and the mass of the solution.   If one bubble is much larger than the other, the larger one will expand while the smaller one contracts.  As the initial size of the smaller bubble goes to zero the solution becomes arbitrarily light.   Further, for low energy bubbles the curvature is large only in a small region and is vanishingly small elsewhere.  This suggests that asymptotically flat five dimensional space may have an instability, although one which requires some initial energy to stimulate.  This amount of energy may, apparently, be arbitrarily small and physically one can only specify the energy of a system only up to some finite resolution.

Explicitly the simplest form of the solutions is
\be
ds^2 = \frac{{r_0}^2}{\omega (\bar{x} + \bar{y} - \bar{x} \bar{y})^2} \Big[ A(\bar{x}) \Big[\frac{4 P(\bar{y})}{B(\bar{y})} d{\bar{\psi}}^2 + \frac{B(\bar{y})}{P(\bar{y})} d{\bar{y}}^2 \Big] 
\ee
$$
+ B(\bar{y}) \Big[ \frac{A(\bar{x})}{R(\bar{x})} d{\bar{x}}^2 + \frac{4 R(\bar{x})}{A(\bar{x})} d\bar{\phi}^2 \Big] \Big]
$$
where
\be
A(\bar{x}) = \Big[ 1 - \Big(1 - \sqrt{1 - \omega}\Big) \bar{x} \Big]^2
\ee
\be
B(\bar{y}) = \Big[ 1 - \Big(1 - \sqrt{\omega}\Big) \bar{y} \Big]^2
\ee
\be
P(\bar{y}) = (1- \bar{y}) \bar{y} (1 - (1 - \omega) \bar{y})
\ee
and
\be
R(\bar{x}) = (1- \bar{x}) \bar{x} (1 - \omega \bar{x})
\ee
The angles $\bar{\psi}$ and $\bar{\phi}$ have period $2\pi$, $0 \leq \bar{x} \leq 1$, $0 \leq \bar{y} \leq 1$ and $0 < \omega < 1$.  One has bubbles at $\bar{y} = 1$ with area $4 \pi {r_0}^2$ and at $\bar{x} = 1$ with area $4\pi {r'_0}^2 = 4 \pi {r_0}^2 (1 - \omega)/\omega$.  Spatial infinity is at $\bar{x} = \bar{y} = 0$.  This coordinate system is illustrated in Figure 2.
\begin{figure}
\centering

	\includegraphics[scale= 0.7]{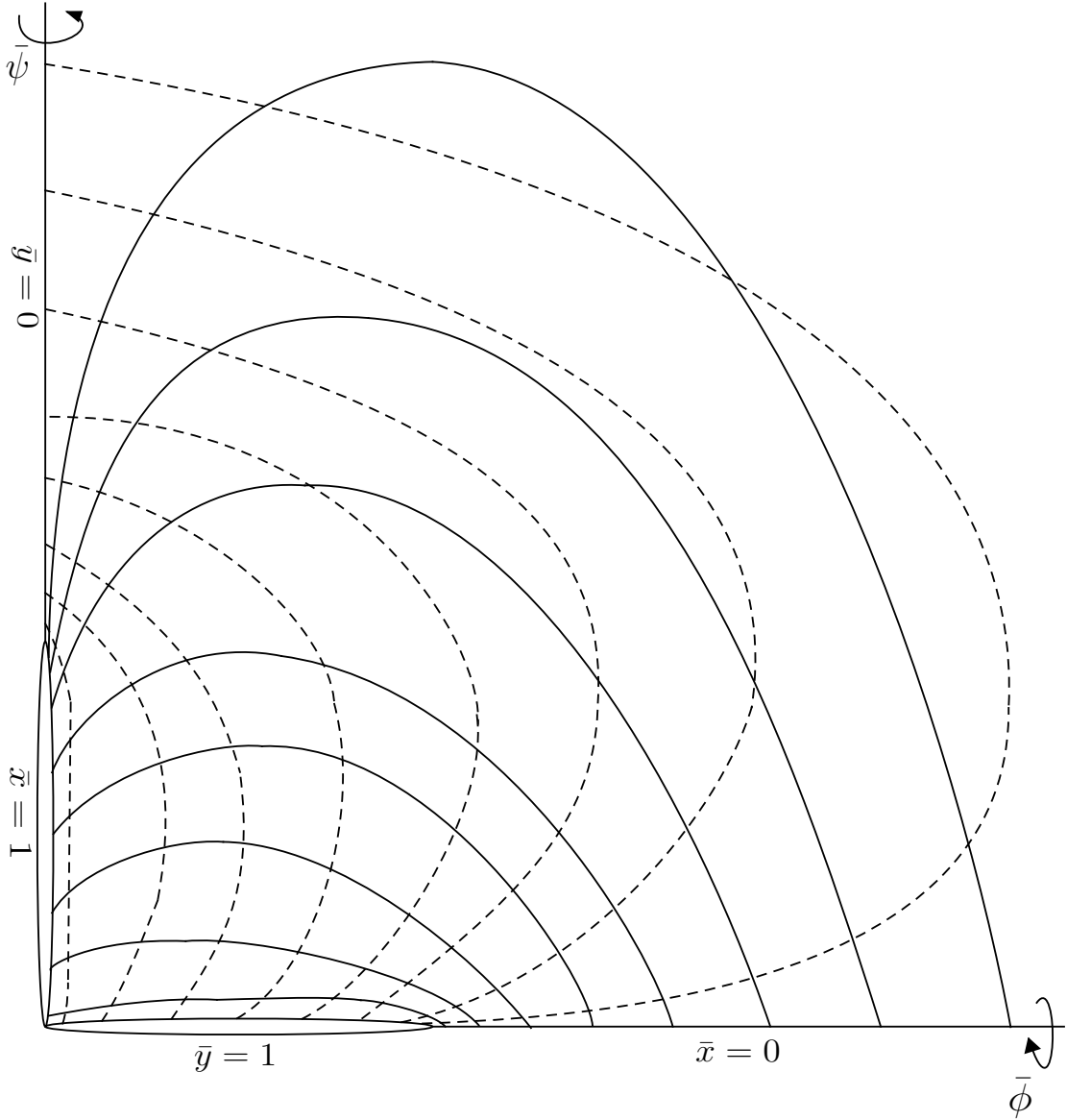}
	\caption{Lines of constant $\bar{y}$ (solid) and constant $\bar{x}$ (dashed).  $\bar{y} = 0$ and $\bar{x} = 0$ are axis for $\bar{\psi}$ and $\bar{\phi}$ respectively and bubbles are formed at $\bar{y} = 1$ as $\bar{\psi}$ pinches off and at $\bar{x} = 1$  as $\bar{\phi}$ pinches off.}
	
	\end{figure}

Perhaps the most important open question regarding these solutions is the time evolution of this initial data.  This must be addressed, apparently, via numerical techniques, although I have argued physically bubbles which start expanding should continue to do so.  It would also be very interesting to numerically study the existence of apparent horizons and the formation of black holes via the collapse of bubbles.

Even in five dimensions, it is possible there are many more bubbles than those described here.   One possible route of advance is to allow a more generic form of the original ansatz (\ref{anstz}) and, in particular,  non-diagonal terms.   There may also be single asymptotically flat bubbles, analogous to those found by Ross \cite{RossAdSBubbles} in AdS.  It would also be interesting to find static solutions and in particular static solutions which, like these bubbles, can be made arbitrarily small while restricting the region of high curvature to a still smaller region.

Further, one can imagine embedding bubbles in any locally flat system provided they are much smaller than scale of the original system.   In particular, there ought to exist AdS versions of these bubbles.  The presence of a cosmological constant prevents the ansatz (\ref{anstz}) from producing a solution with the correct asymptotics and so one must find another way to find these bubbles.  In addition to the obvious interest vis-a-vis AdS-CFT, one might expect that AdS should prevent such bubbles from expanding indefinitely.  One also ought to be able to find such bubbles in compactifications provided the compactification scale is much greater than the bubble size.   Even if these bubbles do expand indefinitely in asymptotically flat space, one might hope to find a compactification where they do not.   The prospect of an instability or low energy signals for generic compactification manifolds (at least for those which are locally geometric) clearly deserves attention.   For braneworld scenarios, it is unclear whether or not this process is significant.   If a bubble nucleated intersecting a brane, an observer on that brane would presumably see the same instability as that described here.  It is not entirely clear, however, whether the matter of brane might prevent such nucleation.  If a bubble nucleated away from the brane, even if it expanded outward forever it would presumably just push the brane along with it.  Whether the consequences to an observer on the brane would be severe, or even significant, is unknown.

Whether or not these bubbles indicate a true instability of higher dimensions remains an open question.  This question is sensitively dependent on whether there are any bubbles which expand for a significant period of time--if all expanding bubbles immediately stop expanding and collapse there would not appear to be any cause for concern.   In the absence of anything preventing the formation of these bubbles, one would expect their formation would be compulsory.  There are no conserved charges preventing the formation of these bubbles and at least at present there does not appear to be any topological argument to rule out their formation.  While there is no obvious obstruction, it has not yet been shown these bubbles are cobordant to flat space or that a suitable spin structure exists on these manifolds.  Such considerations could conceivably act to stabilize higher dimensions.

I have not discussed the quantum mechanical process of nucleating such bubbles.   Finding an instanton to produce these or other similar bubbles would be quite interesting but represents a significant challenge.  However, it is difficult to see how the production of these bubbles could be generically small.  Even if the relevant probability contained a Planck level supression, as Witten's bubble does, one could simply examine bubbles an order of magnitude larger than the Planck scale.  Further, from a path integral perspective, there are presumably many quantum paths of action of order $\hbar$ producing bubbles several times the Planck size which can then expand outwards rapidly.  Hence a process of quantum mechanical production and classical expansion would seem to produce many such bubbles even if no instanton could be found.   On the other hand, one might interpret these observations as evidence that transitions in quantum gravity must be more restricted than one normally expects.

It seems extremely likely that there are higher dimensional analogues of these bubbles.  As with black rings, however, it is not entirely clear with what ansatz one ought to begin.   Also as with black rings, one might imagine the class of solutions is much larger in higher dimensions; for example, one can imagine a sphere or even more generic manifold pinching off instead of a circle.  The dynamical behavior of such bubbles, and in particular expansion or contraction, is also of interest.  However, if these five dimensional bubbles expand indefinitely it seems likely there will be bubbles in higher dimensions that also do so.  The reason is simply that one does not expect an instability in a dimensionally reduced description to disappear in the higher dimensional theory.

It would be interesting to better understand quantum mechanical effects on these backgrounds.  While the curvature is low in most of the spacetime, Mathur and collaborators \cite{Mathur} have suggested low curvature regions can obtain large stringy corrections.  To substantially alter the results in this case, however, such effects would have to extend far outside of the region of any semiclassical horizon.  The solutions do appear to be safe from closed string tachyons \cite{Clstachys}; any time one has an $S^1$ of the string scale the circle pinches off within that same scale.

\vskip 1cm
\centerline{\bf Acknowledgments}
\vskip .5cm

It is a pleasure to thank  D. Marolf,  S. Ross and especially G. T. Horowitz for useful discussions.   This work was supported by grant NSF-PHY-0555669.

 \end{document}